\documentclass[twocolumn,aps,prb,superscriptaddress]{revtex4}
\usepackage{amssymb}
\usepackage{graphicx}
\usepackage{dcolumn}
\usepackage{bm}
\usepackage{amsmath}

\begin{document}

\title{Diagnosis of bulk phase diagram of non-reciprocal topological
lattices by impurity modes}
\author{Yanxia Liu}
\affiliation{Beijing National Laboratory for Condensed Matter Physics, Institute of
Physics, Chinese Academy of Sciences, Beijing 100190, China}
\author{Shu Chen}
\email{schen@iphy.ac.cn}
\affiliation{Beijing National Laboratory for Condensed Matter Physics, Institute of
Physics, Chinese Academy of Sciences, Beijing 100190, China}
\affiliation{School of Physical Sciences, University of Chinese Academy of Sciences,
Beijing 100049, China}
\affiliation{Yangtze River Delta Physics Research Center, Liyang, Jiangsu 213300, China}

\begin{abstract}
We study the single impurity problem in the non-Hermitian lattice described
by the non-reciprocal Su-Schrieffer-Heeger model and obtain the phase
diagram of localized bound states induced by the impurity. The existence of
analytical results permits us to determine the phase boundaries exactly,
which separate regions with different number of bound states. Particularly,
in the limit of strong impurity potential, we find that the phase boundaries
of mid-gap bound states are identical to the boundaries of topological
phase diagrams of the bulk system in the absence of impurity. The existence
of correspondence between mid-gap impurity modes and bulk phases indicate
that we are able to diagnose the bulk phase diagram of the non-reciprocal
topological model by its impurity modes.
\end{abstract}

\maketitle


\section{Introduction}
Non-Hermitian systems have attracted much attention in the past years \cite%
{Moiseyev} as some open quantum and optic systems can be effectively modeled
by Hamiltonians with non-Hermitian terms \cite%
{Ozawa,Lu,Guo,Peng,Zhen,Doppler,Ding,Midya,Xu,Ghatak,Alvarez1}. Due to the lack of Hermitian
restriction, the non-Hermitian systems posses more types of fundamental
nonspatial symmetries than their Hermitian counterpart and thus their
topological classification goes beyond the standard ten classes \cite%
{Gong,LiuCH1,Sato,Zhou,Lieu-BDG,Karabata,SongZhi}. It was shown that the non-Hermitian systems may
exhibit rich exotic phenomena without Hermitian counterparts \cite{Hatano,HatanoPRB,Levitov,Esaki,ZhuBG,Yuce2015,XuY,Leykam,ShenH,TELee,Yin,Lieu,JiangHui,Yao,Alvarez,Xiong,Kunst}, e.g., the appearance of novel topological invariants \cite{TELee,Yin,Lieu,JiangHui,ShenH,Leykam}, the existence
of non-Hermitian skin effect (NHSE) and breakdown of usual bulk-boundary
correspondence in the non-reciprocal lattices \cite%
{Yao,Alvarez,Xiong,Gong,Kunst,Yao1,Jin,Lee,Kawabata,Lee1,WangZhong2019,Herviou,Kou,Rui,Xue,Longhi,JiangHui2019,Turker}. A generalized
Brillouin zone was proposed to recover the correspondence between the
winding of complex energy with periodic boundary and the existence of skin
modes with open boundary \cite{Yokomizo,Fang,HuJP,Okuma}. The interplay of skin effect
and disorder was also studied recently \cite{JiangHui-PRB}.

In general, nontrivial topological properties of bulk systems can be
detected by the behaviors of defects, such as impurities, edges and
dislocations \cite{Kane,LiuCH2,Slager2015,ShenSQ,Kimme,Lang2014,Borgnia}.
A domain wall configuration is a typical topological defect, which is
exponentially localized at the interface between topologically different
phases.
The open boundary can be viewed as a special domain wall, which separates
the bulk state from the vacuum, and the edge state emerges at the boundary
if the bulk state is topologically nontrivial. The domain wall in some
non-Hermitian topological systems was studied recently and the non-Hermitian
effect on the defect state was discussed \cite{Schomerus,Malzard,Yuce,Lang,Deng,WuYJ,Ezawa,Longhi2014}. While
previous works mainly focused on the topological defect on non-Hermitian
lattices with spatially distributed gain and loss, the impurity problem in
non-reciprocal lattices was rarely addressed. A prominent feature induced by
nonreciprocal hopping processes is the emergence of NHSE under open boundary
condition (OBC). The breakdown of bulk-edge correspondence in nonreciprocal
lattices suggests the sensitivity of boundary condition for these systems \cite{Borgnia}.
For a single impurity in a one-dimensional (1D) Hermitian lattice, the impurity always induces
gap bound states, which are exponentially localized at the impurity, and
meanwhile the forward and backward scattering amplitude of the continuous states can be tuned by
varying the impurity strength. Particularly, if the strength goes to
infinity, the forward scattering is forbidden, and the impurity potential
can be effectively treated as a hard-wall boundary (or an OBC). When the
impurity strength increases from zero to infinity, a ring lattice with a
single impurity can continuously interpolate the system with periodic
boundary condition (PBC) to OBC, excluding the gap bound states.
However, such an interpolation is broken down for the single impurity problem in the non-reciprocal lattices. The spectrum of continuous states in the limit of infinity strong impurity potential is obviously different from the spectrum of the open boundary lattice.
Therefore, it is very interesting to explore how the impurity potential affects the properties of bound and continuous states in the non-reciprocal topological lattices.

In this work, we focus on a single impurity problem in a non-Hermitian
lattice described by a non-reciprocal Su-Schrieffer-Heeger (SSH) model,
which has been shown to exhibit NHSE and breakdown of bulk-boundary
correspondence. To begin with, we shall first study the single impurity problem in a simple one-band non-reciprocal lattice, which can be solved analytically.
Then we analytically solve the impurity problem in the non-reciprocal SSH model and obtain the
phase diagram of localized bound states induced by the impurity, in which
different phases are distinguished by numbers of bound states. In the limit
of large impurity strength, the energy of mid-gap bound state approaches
zero and we find that the phase boundaries of mid-gap zero-mode state are
identical to the boundaries of bulk topological phase diagrams, indicating
that we can detect the bulk phase boundaries of the non-reciprocal
topological model by its impurity modes.

The paper is organized as follows. In Sec. II, we firstly study the single impurity problem in a one-band non-reciprocal chain. With the help of the Green's function
method, we can analytically determine the phase diagram for the formation of
bound state induced by the impurity.  In Sec. III, we study the single impurity problem in the non-reciprocal SSH model, which can be analytically solved by mapping the problem to the impurity problem studied in the previous section.  We summarize our results in Sec. IV.

\section{Single impurity problem in a one-band non-reciprocal lattice}
Firstly, we study a single impurity problem in a simple one-band
non-reciprocal lattice, which is described by%
\begin{equation}
H=H_{0}+V_0 \left\vert 0\right\rangle \left\langle 0\right\vert ,
\label{HNV}
\end{equation}%
where $V_0$ is the strength of the impurity potential on site $0$ and $H_{0}$
is given
\begin{equation}
H_{0}=\sum_{n}\left( t_{L}\left\vert n\right\rangle \left\langle
n+1\right\vert +t_{R}\left\vert n+1\right\rangle \left\langle n\right\vert
\right) ,  \label{H0}
\end{equation}%
where the periodic boundary condition is considered, $t_{R(L)}$ denotes the right (left)-hopping amplitude
which can be parameterized as $t_{L}=te^{-g}$ and $t_{R}=te^{g}$ with real $%
t $ and $g$. For convenience, we take $t=1$. The model (\ref{H0}) is also known as the Hatano-Nelson model \cite{Hatano} without disorder. The non-Hermiticity is induced by the asymmetric hopping processes $\left( g\neq 0\right) $. Under the periodic boundary condition, the Hamiltonian in the
momentum space is given by
\begin{equation}
H_{0} = \sum_k h_{0}(k)\left\vert k\right\rangle \left\langle k\right\vert ,
\end{equation}%
with
\begin{equation}
h_{0}(k) = 2t \cos \left( k+ig\right).
\end{equation}%
As the spectrum is complex, one can define a winding number associated with
the complex eigenvalue as
\begin{equation}
w=\int_{-\pi }^{\pi }\frac{dk}{2\pi i}\partial _{k}\ln h_{0}\left( k\right),
\end{equation}%
which gives rise to
\begin{equation*}
w =\left\{
\begin{array}{cc}
1, & g>0 \\
-1, & g<0%
\end{array}%
\right. .
\end{equation*}%
The phase with winding number $w =1$ or $-1$ belongs to topologically
different phase and the transition point locates at $g=0$, at which the
Hamiltonian reduces to a Hermitian one.

\begin{figure}[tbp]
\includegraphics[width=0.5\textwidth]{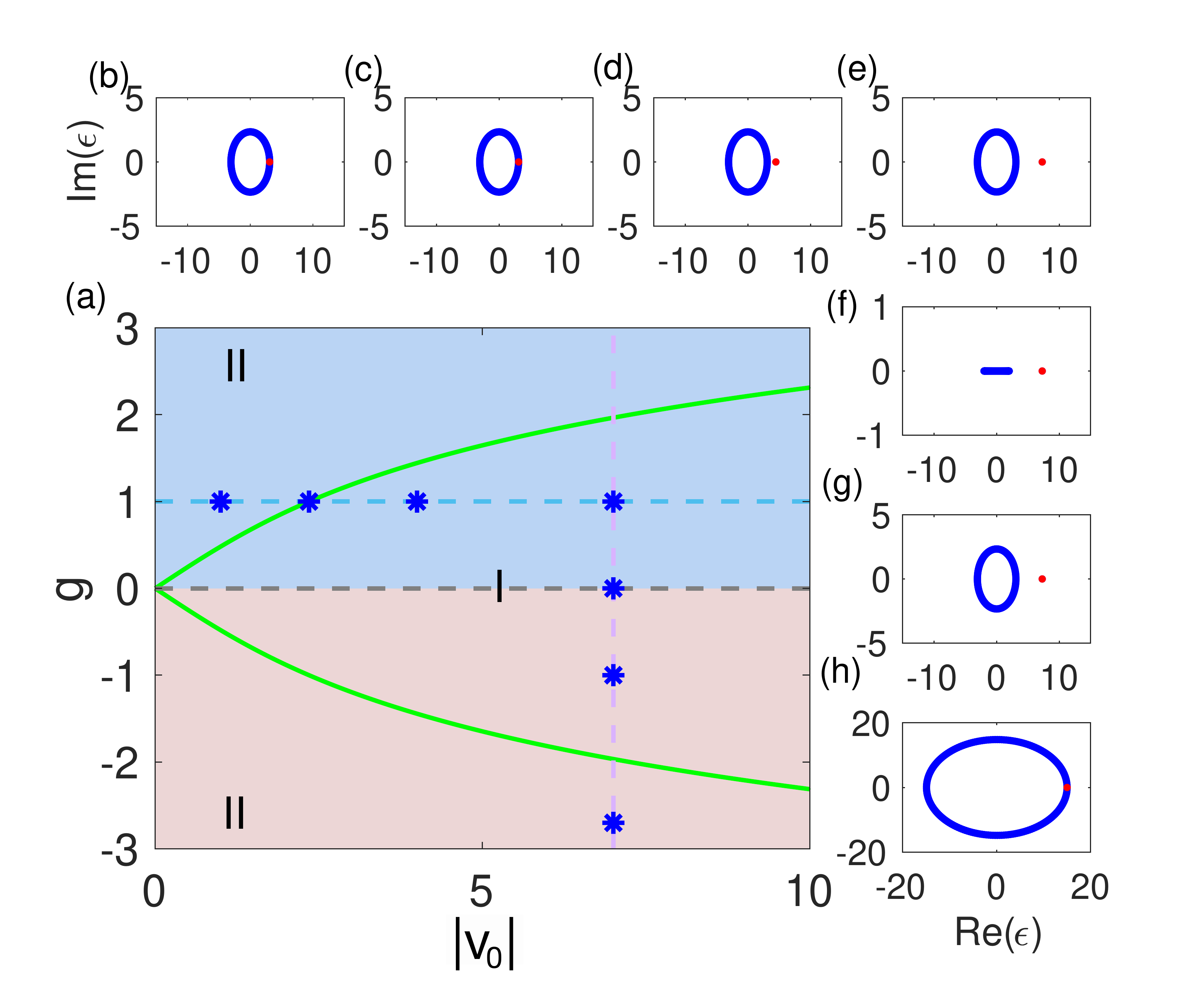}
\caption{(a) Phase diagram for localized bound state induced by the impurity in the 1D non-reciprocal lattice. In the regions (I), there is one bound state, and in the
region (II), there is no bound state. (b)-(e) The complex-energy spectra
with $g=1$ for $v_0=1,2.35,4,7$.(f)-(h) The complex-energy spectra with $v_0=7$
for $g=0,-1,-2.7$. }
\label{fig1}
\end{figure}

With the help of
Green's function, the wavefunction of the Hamiltonian can be written as
\begin{equation}
\psi \left( n\right) =\frac{V_0 \delta _{n,0}}{E-H_{0}}\psi \left( n\right) .
\label{wf}
\end{equation}%
where $E$ is the eigenvalue of the system.
Making a Fourier transformation $\phi \left( k\right) =\sum_{n}e^{-ikn}\psi
\left( n\right) $ and using Eq.(\ref{wf}), we get
\begin{equation*}
\phi \left( k\right) =\frac{V_0 \psi \left( 0\right) }{E-2t\cos \left(
k+ig\right) }.
\end{equation*}%
Then the inverse Fourier transformation of the above equation gives rise to%
\begin{eqnarray*}
\psi \left( n\right)  &=&\int_{-\pi }^{\pi }\frac{dk}{2\pi }e^{ikx}\phi
\left( k\right)  \\
&=&\int_{-\pi }^{\pi }\frac{dk}{2\pi i}\frac{e^{ikx}v_{0}\psi \left(
0\right) }{\epsilon -2\cos \left( k+ig\right) }
\end{eqnarray*}%
Let $y=e^{ik}$, $v_{0}=V_0/t$ and $\epsilon =E/t$, and we can get%
\begin{equation}
\psi \left( n\right) =\oint \frac{dy}{2\pi e^{-g}i}\frac{v_{0}\psi \left(
0\right) }{\left( y_{2}-y_{1}\right) }\left( \frac{y^{n}}{y-y_{1}}-\frac{%
y^{n}}{y-y_{2}}\right) ,  \label{wf1}
\end{equation}%
where the integral is along a unit circle around the original point and
\begin{equation*}
y_{1,2}(\epsilon )=e^{g}\frac{\epsilon \pm \sqrt{\epsilon ^{2}-4}}{2}.
\end{equation*}%
When $n=0$, Eq. $\left( \ref{wf1}\right) $ becomes
\begin{equation}
-\frac{2\pi ie^{g}}{v_{0}}=\oint dy\frac{1}{\left( y_{2}-y_{1}\right) }%
\left( \frac{1}{y-y_{1}}-\frac{1}{y-y_{2}}\right) .  \label{boundeq}
\end{equation}%
The right part of the above equation can be integrated by using the residue
theorem, and the equation has solution only the following condition
\begin{equation}
f\left( v_{0}\right) <e^{-\left\vert g\right\vert }  \label{tran}
\end{equation}%
is fulfilled, where
\begin{equation*}
f\left( v_{0}\right) =\frac{\sqrt{v_{0}^{2}+4}-\left\vert v_{0}\right\vert }{%
2}
\end{equation*}%
with $v_0=V_0/t$. The solution of Eq.(\ref{boundeq}) gives the eigenenergy of
the bound state
\begin{equation}
\epsilon =\text{sgn}(v_{0})\sqrt{v_{0}^{2}+4},  \label{energy1}
\end{equation}%
which is real and irrelevant to the parameter $g$.

By using
\begin{equation}
f\left( v_{0}\right) =e^{-\left\vert g\right\vert },
\end{equation}%
we can determine the phase boundary and get the phase diagram for the
localized bound state induced by the impurity, which is shown in Fig.\ref{fig5}(a). In
the region I, there exists a localized bound state, whereas in the region
II, no bound state exists. In Fig.\ref{fig1}(b)-(h), we display the
distribution of energy spectrum for systems with various parameters. For the
system in the region II, as shown in Fig.\ref{fig1}(b) and (h), only
continuous states exist and their eigenvalues form a close circle in the
complex plane. For the system in the region I, as shown in Fig.\ref{fig5}%
(d), (e) and (g), there exists a bound state with a real eigenvalue
distributed in the real axis outside the circle formed by the continuous
states. When $g=0$, the system reduces to the Hermitian case, as shown in
Fig.\ref{fig1}(f). The bound state vanishes and the corresponding eigenvalue
merges into the circle distribution of continuous states when $g$ locates at
the phase boundary, as shown in Fig.\ref{fig1}(c).  From Fig.\ref{fig1}(b)-(e),
we can see that the complex eigenvalues of continuous states almost keep invariant with the increase in $v_0$.
The spectrum of continuous states has no essential change even when $v_0 \rightarrow \infty$,
which is obviously different from the real spectra of the open boundary system.
\begin{figure}[tbp]
\includegraphics[width=0.5\textwidth]{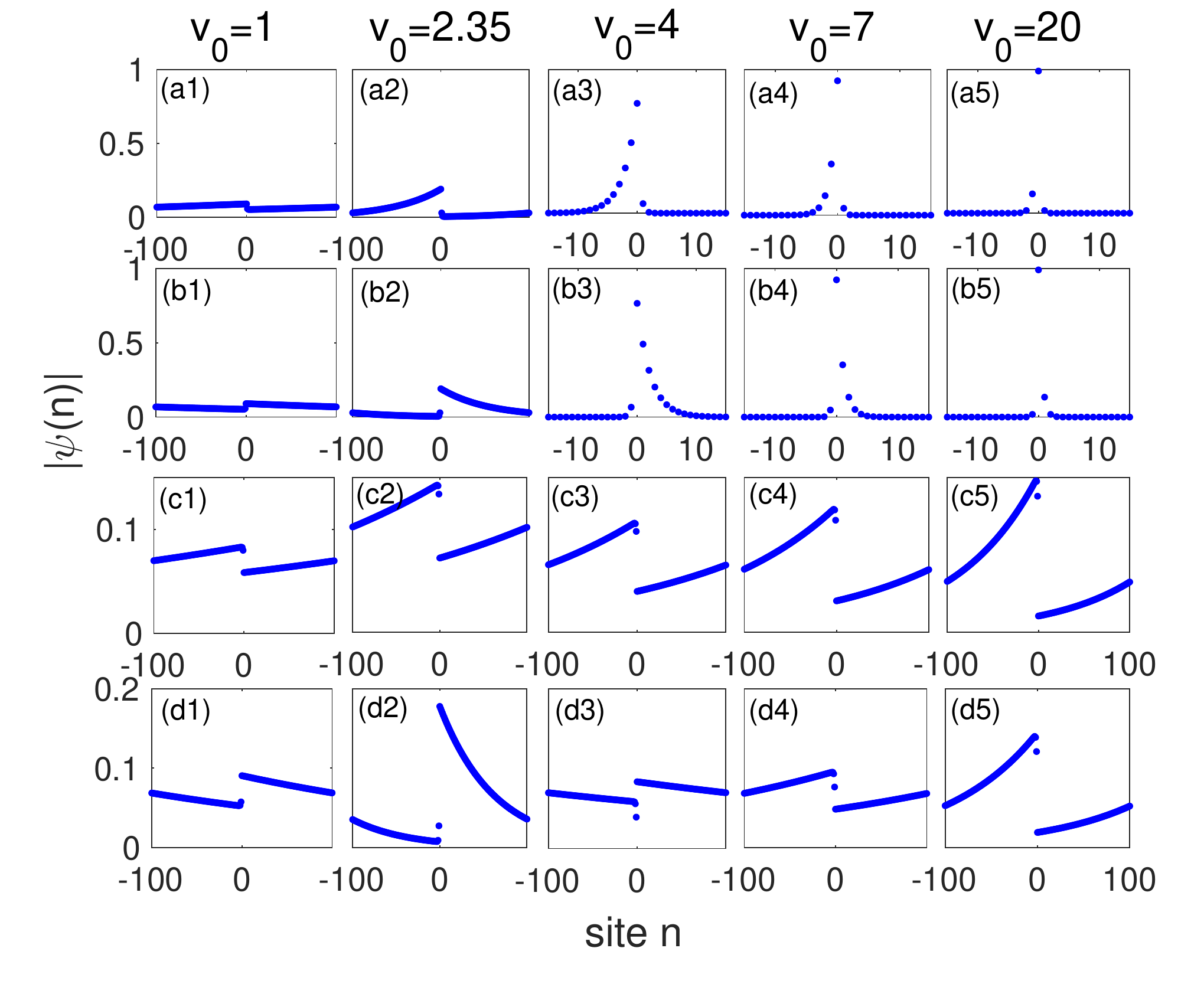}
\caption{The distributions of wavefunctions for the impurity model in the one-band non-reciprocal lattice.
The columns represent results for five parameters $v_0=1$, $2.35$, $4$, $7$ and $20$, respectively.
(a1)-(a5) The distributions for the wavefunctions corresponding  to the bound state with $g=-1$. (b1)-(b5) The distributions
for the wavefunctions corresponding  to the bound state with $g=1$. (c1)-(c5) The distributions of the continuous states with
minimum real part of eigenenergies for the system with $g=1$
(d1)-(d5) The distributions of the continuous states
states with maximum real part of eigenenergies for system with $g=1$.}
\label{fig2}
\end{figure}
\begin{figure}[tbp]
\includegraphics[width=0.5\textwidth]{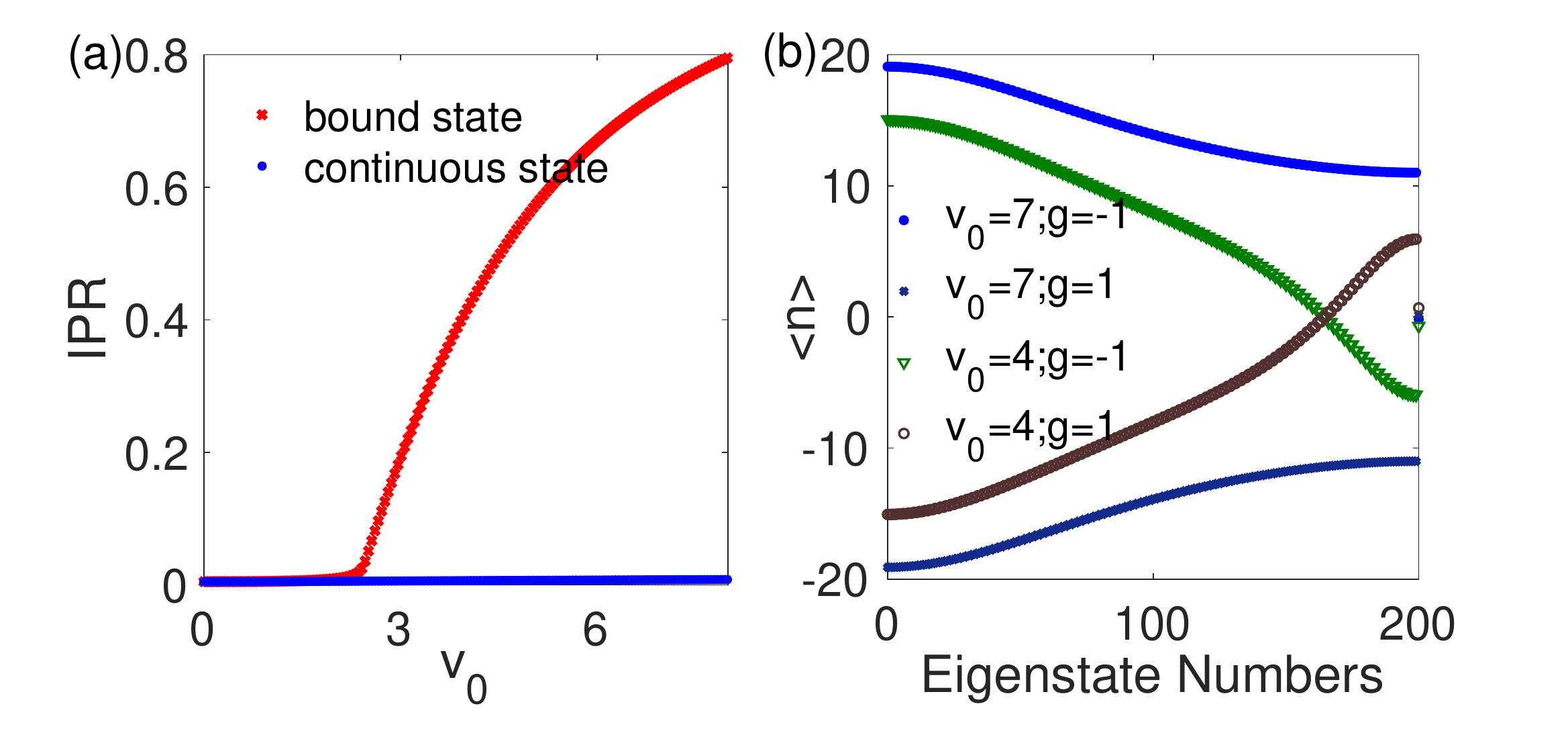}
\caption{(a) The IPR for the wavefunction corresponding to the bound state and one of the continuous state
versus $v_0$  for the system with $g=1$. (b) The mean position of all states for systems with $v_0=4$, $7$ and
$g=\pm1$. }
\label{fig3}
\end{figure}

The corresponding wavefunction of the bound state can be written as%
\begin{equation}
\psi _{b}\left( n\right) \propto \left\{
\begin{array}{cc}
\left[ \text{sgn}(v_{0})e^{g}f\left( v_{0}\right) \right] ^{n} & n>0, \\
\left[ \text{sgn}(v_{0})e^{-g}f\left( v_{0}\right) \right] ^{\left\vert
n\right\vert } & n<0,%
\end{array}%
\right.   \label{boundstateB}
\end{equation}%
where $\text{sgn}(v_{0})$ is the sign function. When $g=0$, the bound state
decays exponentially and distributes symmetrically around $n=0$.
When $g<0$, the wavefunction $\psi _{b}(n)$
distributes asymmetrically and decays more quickly in the region $n>0$ than
in the region $n<0$, as shown in Fig.\ref{fig2}(a3)-(a5). When $g>0$, the
wavefunction decays more quickly in the region $n<0$ than in the region $n>0$%
, as shown in Fig.\ref{fig2}(b3)-(b5).
From Eq.(\ref{boundstateB}), we can also see that the existence of bound state enforces the constraint condition
$e^{|g|}f(v_{0})<1$, otherwise Eq.(\ref{boundstateB}) is not a bound
state located around $n=0$.
 If $f(v_{0})\geqslant e^{-|g|}$, the bound state vanishes and becomes a continuous state, as shown in Fig.%
\ref{fig2}(a1),(a2),(b1) and (b2). For the extended states as shown in the
third and fourth rows of Fig.\ref{fig2}, no obvious change in the
distribution of wavefunction is observed when $v_0$ crosses the transition
point $v_0=2.35$. To see clearly the transition between the localized (bound)
state and extended (continuous) state, we calculate the inverse
participation ratio (IPR) of the eigenstate $\left\vert \psi \right\rangle $
\begin{equation*}
\text{IPR}=\frac{\sum_{n}\left\vert \left\langle n|\psi \right\rangle
\right\vert ^{4}}{\left( \left\langle \psi |\psi \right\rangle \right) ^{2}}.
\end{equation*}%
While $\text{IPR}=1$ corresponds to a perfectly localized state, $\text{IPR}=1/N
$ corresponds to a fully extended state. We display the IPR for the state
corresponding to the bound state versus $g$ in Fig. \ref{fig3}(a), which
indicates clearly a transition from an extended to localized state around $%
v_0=2.35$. As a comparison, we also show the IPR for a
continuous state in the same figure, which always approaches zero in the
whole region of $v_0$.

In Fig.\ref{fig2}, we have shown that the distribution of wavefunction is
asymmetrical with $g\neq0$. To characterize the asymmetry,
we can calculate the mean position $\left\langle n\right\rangle
=\sum_{n}\left\langle \psi \right\vert n\left\vert \psi \right\rangle $.
While $\left\langle n\right\rangle =0$ corresponds to a symmetrical
distribution, $\left\langle n\right\rangle >0$ ($<0$) indicates that the
wavefunction more likely distributes on the right (left) region of the
impurity. The distribution tendency of the bound state is solely determined
by $g>0$ or $g<0$. For large $v_0$, the distribution of extended state
also depends on the sign of $g$. For example, the wave function tends to distribute on the
left  region of the impurity with $g>0$ for the large $v$, as shown in Fig.
\ref{fig2} (c4),(c5),(d4) and (d5). However the extended state can distribute on either the
right or left region of the impurity for a given $g$ depending on the choice
of state for small $v_0$, as shown in Fig.\ref{fig2}(b1)-(d1), (b2)-(d2) and (c3)(d3), respectively.
To see it clearly, we display the mean position $\left\langle n\right\rangle
$ for all states of the system with $v_{0}=4,7$ and $g=\pm 1$ in Fig.\ref%
{fig3}(b).
The eigenstates are sorted by the real part of the corresponding
eigenenergies. We can see that the mean positions of eigenstates with $g=1$
and $g=-1$ are symmetric about $\left\langle n\right\rangle =0$.

\section{Single impurity problem in the  non-reciprocal SSH model}
\subsection{Model and bulk phase diagram}
\begin{figure}[tbp]
\includegraphics[width=0.5\textwidth]{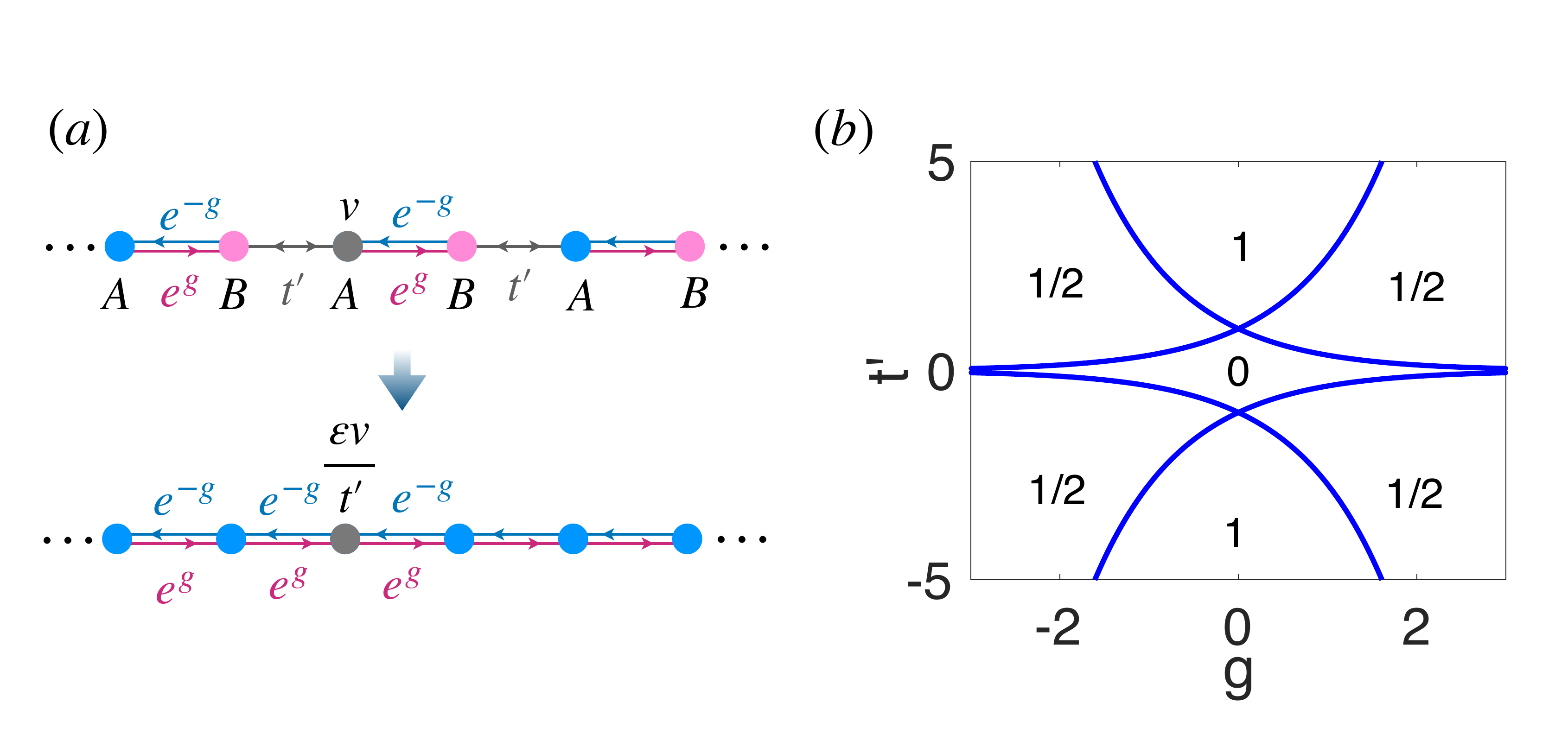}
\caption{(a) Schematic diagram of the non-reciprocal SSH model with an impurity.
This model can be mapped into a one-band non-reciprocal lattice model with an
impurity. (b) The phase diagram of the non-reciprocal SSH model with topologically different phases characterized by the topological invariant
$\nu_s=0$, $1/2$ and $1$.}
\label{fig4}
\end{figure}

We consider the single impurity problem in a non-reciprocal SSH model with
different hopping amplitudes along the right and left hopping directions in
the unit cell, as schematically displayed in Fig.\ref{fig4}(a). The Hamiltonian is
described by
\begin{equation}
\mathcal{H}=\mathcal{H}_{NSSH}+ V \left\vert 0,A\right\rangle \left\langle
0,A\right\vert ,
\end{equation}%
where $V$ is the strength of the impurity potential and
\begin{eqnarray}
\mathcal{H}_{NSSH} &=&\sum_{n}\left[ t (e^{-g}\left\vert n,A\right\rangle
\left\langle n,B\right\vert +e^{g}\left\vert n,B\right\rangle \left\langle
n,A\right\vert ) + \right.  \notag \\
&& \left. t^{\prime } ( \left\vert n,B\right\rangle \left\langle
n+1,A\right\vert + \left\vert n+1,A\right\rangle \left\langle n,B\right\vert
) \right].
\end{eqnarray}%
Here $te^{\pm g}$ is the right (left) intra-hopping amplitude, $t^{\prime }$
is the inter-hopping amplitude, $A$ ($B$) represents the sublattice labels
and $n$ indicates the $n$-th cell of the lattice. For convenience, we also
set $v=V/t$ and $\varepsilon=E/t$. In the following calculation, we take $%
t=1 $ as the unit of energy.

In the absence of impurity, the non-reciprocal SSH model can be represented
in the momentum space via a Fourier transformation with the Hamiltonian
given by%
\begin{equation*}
\mathcal{H}_{NSSH}\left( k\right) =h_{x}\sigma _{x}+h_{y}\sigma _{y},
\end{equation*}%
where $h_{x}=t^{\prime }\cos k+\cosh g$, $h_{y}=t^{\prime }\sin k-i\sinh g$
and 
$\sigma _{x,y}$ are the Pauli matrices. The phase diagram of the
non-reciprocal SSH model can be obtained by following Refs.\cite{Yin,Yao} with the phase
boundaries determined by
\begin{equation}
t^{\prime }=-e^{\pm g}~~~\text{and}~~~t^{\prime }=e^{\pm g}. \label{condition1}
\end{equation}%
The topologically different phases can be distinguished by the
topological invariant
$ \nu _{s} = \frac{1}{\pi }\int_{-\pi }^{\pi }dk\left\langle \phi
_{\pm}^{L}\right\vert i\partial _{k}\left\vert \phi _{\pm}^{R}\right\rangle$,
which takes%
\begin{equation*}
\nu _{s}=\left\{
\begin{array}{cc}
1 & \text{ }\left\vert t^{\prime }\right\vert >e^{\left\vert g\right\vert }%
\text{ } \\
\frac{1}{2} & e^{-\left\vert g\right\vert }<\left\vert t^{\prime
}\right\vert <e^{\left\vert g\right\vert }\text{ } \\
0 & \left\vert t^{\prime }\right\vert <e^{-\left\vert g\right\vert }%
\end{array}%
\right. ,
\end{equation*}%
where $\left\vert
\phi _{\pm }^{R}\right\rangle$ are the eigenvectors of $\mathcal{H}_{0}\left( k\right) $  and $\left\langle \phi
_{\pm }^{L}\right\vert$ the corresponding left vectors.
The phase diagram is shown in the Fig. \ref{fig4}(b).
\begin{figure}[tbp]
\includegraphics[width=0.5\textwidth]{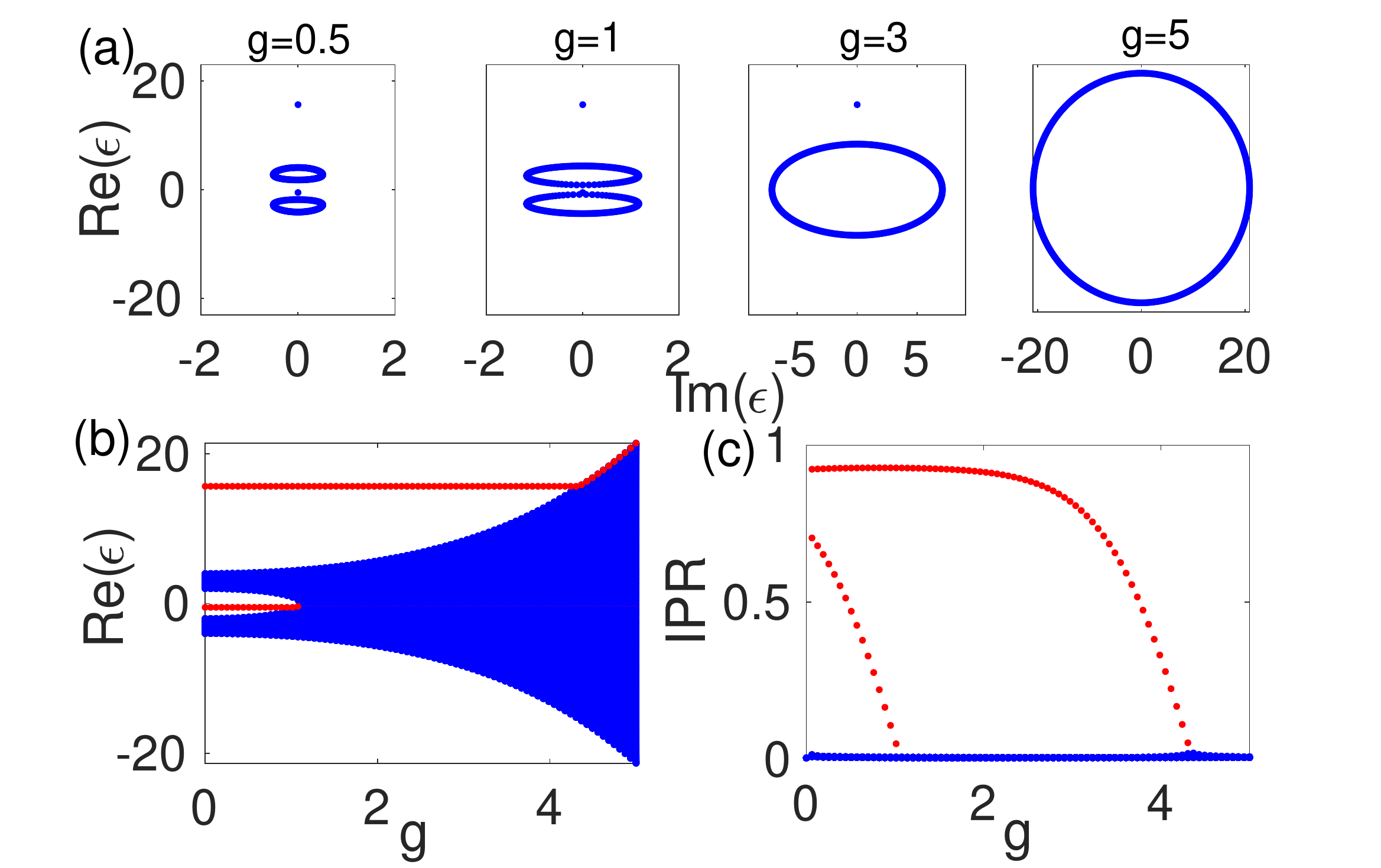}
\caption{(a)The complex-energy spectra for the system with $t^{\prime }=3$,
$v=15$, $g=0.5$, $1$, $3$ and $5$, respectively. (b) The real part of the energy spectra versus $g$ for the system with $t^{\prime }=3$ and $v=15$. (c) IPR for two bound states (red dots) and other continuous states (blue dots).}
\label{fig5}
\end{figure}

\subsection{Solution to the impurity problem}
Now we study the single impurity problem by solving the stationary Schr\"{o}%
dinger equation $\mathcal{H}\left\vert \psi \right\rangle =E\left\vert \psi
\right\rangle $ with $\left\vert \psi \right\rangle =\sum_{n}\left( \psi
_{A}\left( n\right) \left\vert n,A\right\rangle +\psi _{B}\left( n\right)
\right) \left\vert n,B\right\rangle $, which is equivalent to solving the
following recurrence equations%
\begin{eqnarray}
e^{g}\psi _{A}\left( n-1\right) +t^{\prime }\psi _{A}\left( n\right)
=\varepsilon \psi _{B}\left( n-1\right) ,  \label{re1} \\
t^{\prime }\psi _{B}\left( n-1\right) +e^{-g}\psi _{B}\left( n\right) +\delta
_{x,0}v\psi _{A}\left( n\right) =\varepsilon \psi _{A}\left( n\right),
\label{re2}
\end{eqnarray}%
where $\varepsilon$ is the eigenvalue. From Eq.(\ref{re1}), we get
\begin{equation}
\psi _{B}\left( n\right) = \frac{e^{g}}{\varepsilon }\psi _{A}\left(
n\right) + \frac{t^{\prime }}{\varepsilon }\psi _{A}\left( n+1\right) .
\label{re3}
\end{equation}%
By substituting it into Eq.(\ref{re2}), we get%
\begin{eqnarray}
& & e^{g}\psi _{A}\left( n-1\right) +e^{-g}\psi _{A}\left( n+1\right)
+\delta _{x,0}\frac{\varepsilon v}{t^{\prime }}\psi _{A}\left( n\right)
\notag \\
&=& \frac{\varepsilon ^{2}-t^{\prime 2}-1}{t^{\prime }}\psi _{A}\left(
n\right).  \label{Eq-A-lattice}
\end{eqnarray}

It is straightforward to see that the above equation can be mapped into an
impurity problem described by Eq.(\ref{HNV}) in the previous section,
for which the eigenquation $H \left\vert \psi \right\rangle = \epsilon
\left\vert \psi \right\rangle $ with $\left\vert \psi \right\rangle
=\sum_{n} \psi \left( n\right) \left\vert n \right\rangle$ gives rise to
\begin{equation}
e^{g}\psi \left( n-1\right) +e^{-g}\psi \left( n+1\right) +\delta_{x,0} v_0
\psi \left( n\right) = \epsilon \psi \left( n\right).  \label{eigeneq}
\end{equation}
In comparison with Eq.(\ref{Eq-A-lattice}), it is clear that solving Eq.(\ref%
{Eq-A-lattice}) is equivalent to solving Eq.(\ref{eigeneq}) by making the
following substitution
\begin{eqnarray}
& &v_0 = \frac{\varepsilon v}{t^{\prime }},  \label{veff} \\
& &\epsilon = \frac{\varepsilon ^{2}-t^{\prime 2}-1}{t^{\prime }}.
\label{Eeff}
\end{eqnarray}
Following the procedures described in the previous section, we can analytically determine the condition for the
existence of the bound state and get the  analytical solutions
of the bound state.

Now we can derive the condition for the existence of localized bound state
by substituting Eq.(\ref{veff}) into Eq.(\ref{tran}), which gives rise to
\begin{equation}
f\left( \frac{\varepsilon v}{t^{\prime }}\right) <e^{-\left\vert
g\right\vert }.  \label{boundary1}
\end{equation}%
Similarly by substituting Eq.(\ref{veff}) and Eq.(\ref{Eeff}) into Eq.(\ref%
{energy1}), we can determine the bound energy of the impurity in the
non-reciprocal SSH lattice by solving
\begin{equation*}
\left( \varepsilon ^{2}-t^{\prime 2}-1\right) ^{2}=\left( \varepsilon
v\right) ^{2}+4t^{\prime 2} ,
\end{equation*}%
which gives rise to two solutions:
\begin{equation*}
\varepsilon _{\pm }=\text{sgn}(\pm v)\sqrt{\frac{v^{2}+2t^{\prime 2}+2\pm
\sqrt{(4+v^{2})(4t^{\prime 2}+v^{2})}}{2}}.
\end{equation*}%
From the above equation, we can see the bound energies $\varepsilon _{+}$
and $\varepsilon _{-}$ are always real. While $\varepsilon _{+}$ lies above
or below the continuous bands, $\varepsilon _{-}$ lies in the mid-gap
region. In the large $v$ limit, we have
\begin{equation}
\varepsilon _{-} = - \frac{1}{v}|t^{\prime 2}-1|.  \label{expansion}
\end{equation}
For $|t^{\prime }| \neq 1$, the above equation suggests $\varepsilon _{-}$
approaching zero as $v \rightarrow \infty$. While the real bound energies
are independent of the parameter $g$, the complex eigenvalues of the
continuous states rely on $g$. When $g$ keeps increasing and reaches a
critical value, $\varepsilon _{-}$ firstly merges into the continuous band
accompanying with the disappear of the lower bound state, and then $%
\varepsilon _{+}$ merges into the continuous band as $g$ exceeds the second
critical value, as shown in Fig.\ref{fig5}(a) and (b). Such a transition can
be also detected from the change of inverse participation ratio (IPR) for
states corresponding to the bound states as shown in Fig.\ref{fig5}(c). The
IPR of an eigenstate $\left\vert \psi \right\rangle $ is defined by $\text{%
IPR}=\frac{\sum_{n}\left\vert \left\langle n|\psi \right\rangle \right\vert
^{4}}{\left( \left\langle \psi |\psi \right\rangle \right) ^{2}}. $
We can find that the mid-gap bound state vanishes at $%
g=1.07 $, which is close to the the topological phase boundary $g=\ln 3=
1.0986$. The higher bound state vanishes at $g=4.36$, which is same as the boundary condition obtained by Eq.(\ref{boundary1})
with $\varepsilon =\varepsilon _{+}$.
\begin{figure}[tbp]
\includegraphics[width=0.45\textwidth]{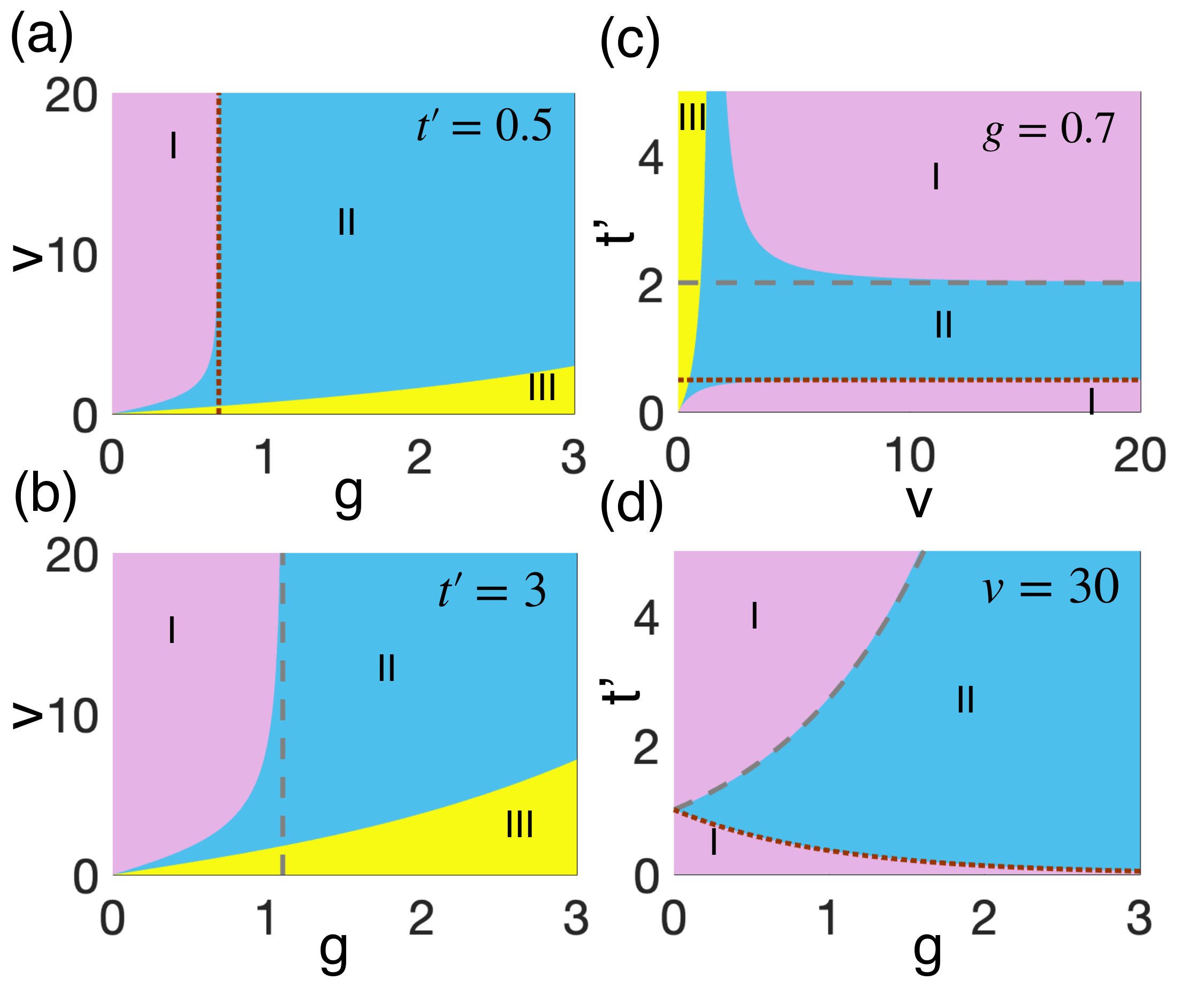}
\caption{ (a)-(d) Phase diagram for bound states induced by a single impurity in the non-reciprocal SSH
model. In the pink (I) and blue (II) regions, there exist two  and one localized  bound states, respectively, and  no bound state exists in the yellow (III) regions. (a) $g$ versus $v$ by fixing $t^{\prime
}=0.5$, (b) $g$ versus $v$
by fixing $t^{\prime }=3$, (c) $v$ versus $t^{\prime }$ by fixing $g=0.7$, (d) $g$ versus $t^{\prime }$ by fixing $v=30$. The
dashed and dotted lines are given by $g=\ln t'$ and $g=-\ln t'$, respectively. }
\label{fig6}
\end{figure}

\subsection{Bound-state phase diagram}
The phase boundaries of bound-state phase diagram can be determined by
\begin{equation}
f\left( \frac{\varepsilon_{-} v}{t^{\prime }}\right) = e^{-\left\vert
g\right\vert }  \label{BSB1}
\end{equation}%
and
\begin{equation}
f\left( \frac{\varepsilon_{+} v}{t^{\prime }}\right) = e^{-\left\vert
g\right\vert }.
\end{equation}%
Different phases in the phase diagram are characterized by the number of bound states. In Fig. \ref{fig3}%
(a)-(d), we display the phase diagrams with different parameters. In regions
I, the parameters satisfy the relations $f\left( \varepsilon
_{\mp}v/t^{\prime }\right) <e^{-\left\vert g\right\vert }$, and there exist
two localized bound states. In regions II, the parameters satisfy the
relations $f\left( \varepsilon _{+}v/t^{\prime }\right) <e^{-\left\vert
g\right\vert } < f\left( \varepsilon _{-}v/t^{\prime }\right) $, there
exists only one bound state. In regions III, no bound state exists as the parameters satisfy the
relation $f\left( \varepsilon _{\pm}v/t^{\prime }\right) > e^{-\left\vert
g\right\vert }$. In Fig. \ref{fig6}(a) and (b),
we display the phase diagram in the parameter space of $v$ versus $g$ with $%
t^{\prime }=0.5$ and $3$, respectively. As $v$ increases and goes toward $%
\infty \,$, the boundary between regions I and region II asymptotically
approaches $g= - \ln 0.5$ (dotted line) and $g= \ln 3$ (dashed line),
respectively. In Fig. \ref{fig3}(c), the phase diagram is plotted for $%
t^{\prime }$ versus $v$ by fixing $g=0.7$. As $v$ tends to $\infty \,$, the
boundaries between regions I and region II approach $t^{\prime }=\exp(0.7)$
(dashed line) and $t^{\prime }=\exp(-0.7)$ (dotted line). Fig. \ref{fig6}(d)
presents the phase diagram in the parameter space of $t^{\prime }$ versus $g$
with $v=30$, which shows the boundaries of the region I and region II almost
coinciding with $t^{\prime }=e^{g}$ and $t^{\prime }=e^{-g}$, respectively.

In the limit of $v \rightarrow \infty$, using the expansion of $%
\varepsilon_{-}$ given by Eq.(\ref{expansion}), we can get
\begin{eqnarray}
f\left( \frac{\varepsilon_{-} v}{t^{\prime }}\right) &=& \frac{1}{2}
(|t^{\prime } + \frac{1}{t^{\prime }}|- |t^{\prime }-\frac{1}{t^{\prime }}|
),  \notag \\
&=& \left\{
\begin{array}{cc}
\frac{1}{|t^{\prime }|} & |t^{\prime }| >1, \\
|t^{\prime }| & |t^{\prime }|<1.%
\end{array}
\right.  \label{limitf}
\end{eqnarray}%
In combination with Eq.(\ref{BSB1}), we get an analytical expression for the
phase boundaries between regions with one and two bound states in the large $%
v$ limit, which reads as
\begin{eqnarray}
&|t^{\prime }|= e^{ \left\vert g\right\vert }, & |t^{\prime }| >1, \\
&|t^{\prime }|=e^{-\left\vert g\right\vert }, & |t^{\prime }|<1 .
\end{eqnarray}%
The analytical results indicate clearly that the boundaries of zero-mode
impurity states are identical to the topological phase boundaries of the
periodical bulk system in the absence of impurity given by Eq.(\ref{condition1}), therefore we can detect the phase boundary of the bulk system
from the boundary of zero-mode bound states induced by a strong impurity
potential.

The distribution of bound state wavefunction on $A$ sites are given by
\begin{equation}
\psi _{A}\left( n\right) =\left\{
\begin{array}{cc}
\left( e^{g}\beta \right) ^{n} & n>0, \\
\left( e^{-g}\beta \right) ^{\left\vert n\right\vert } & n<0,%
\end{array}%
\right.   \label{psiA}
\end{equation}%
where $\beta =\text{sgn}(\frac{\varepsilon _{\pm }v}{t^{\prime }})f\left( \frac{%
\varepsilon _{\pm }v}{t^{\prime }}\right)$.
The distribution of bound state wavefunction on $B$ sites can be obtained
from the recurrence relation $\left( \ref{re3}\right) $, which gives rise to
\begin{equation}
\psi _{B}\left( n\right) =e^{g}\times\left\{
\begin{array}{cc}
\frac{1+t^{\prime }\beta }{\varepsilon }\psi _{A}\left( n\right) & n\geqslant0, \\
\frac{1+t^{\prime }/\beta }{\varepsilon }\psi _{A}\left( n\right) & n<0,%
\end{array}%
\right. .  \label{psiB}
\end{equation}
 In the large $v$
limit, we can get a simple analytical expression for the wavefunction of
mid-gap bound state. When $|t^{\prime }|>1$, we have $\frac{1+t^{\prime
}/\beta }{\varepsilon }=-v$ and $\frac{1+t^{\prime
}\beta }{\varepsilon }=0$, which suggests the wavefuntions mainly
distribute on $B$ sites on the left side of impurity, as the distributions
on $A$ site are greatly suppressed by a factor $1/v$. Similarly,
when $|t^{\prime }|<1$, we have $\frac{1+t^{\prime }/\beta }{\varepsilon }=0$
and $\frac{1+t^{\prime }\beta }{\varepsilon }=-v$,
which suggests the wavefuntions mainly distribute on $B$ sites on the right side
of impurity. In the limit of $v \rightarrow \infty$, the distributions
on $A$ site are completely suppressed, hence the wave function can be written as
\begin{equation*}
\psi \left( n\right)=\psi_{B} \left( n\right) =(-\frac{e^{-g}}{t'})^n,~~~~n<0
\end{equation*}
for $|t'|>1$ and
\begin{equation*}
\psi \left( n\right)=\psi_{B} \left( n\right) =(-e^{g}t')^{|n|},~~~~n>0
\end{equation*}
for $|t'|<1$.
\begin{figure}
\includegraphics[width=0.45\textwidth]{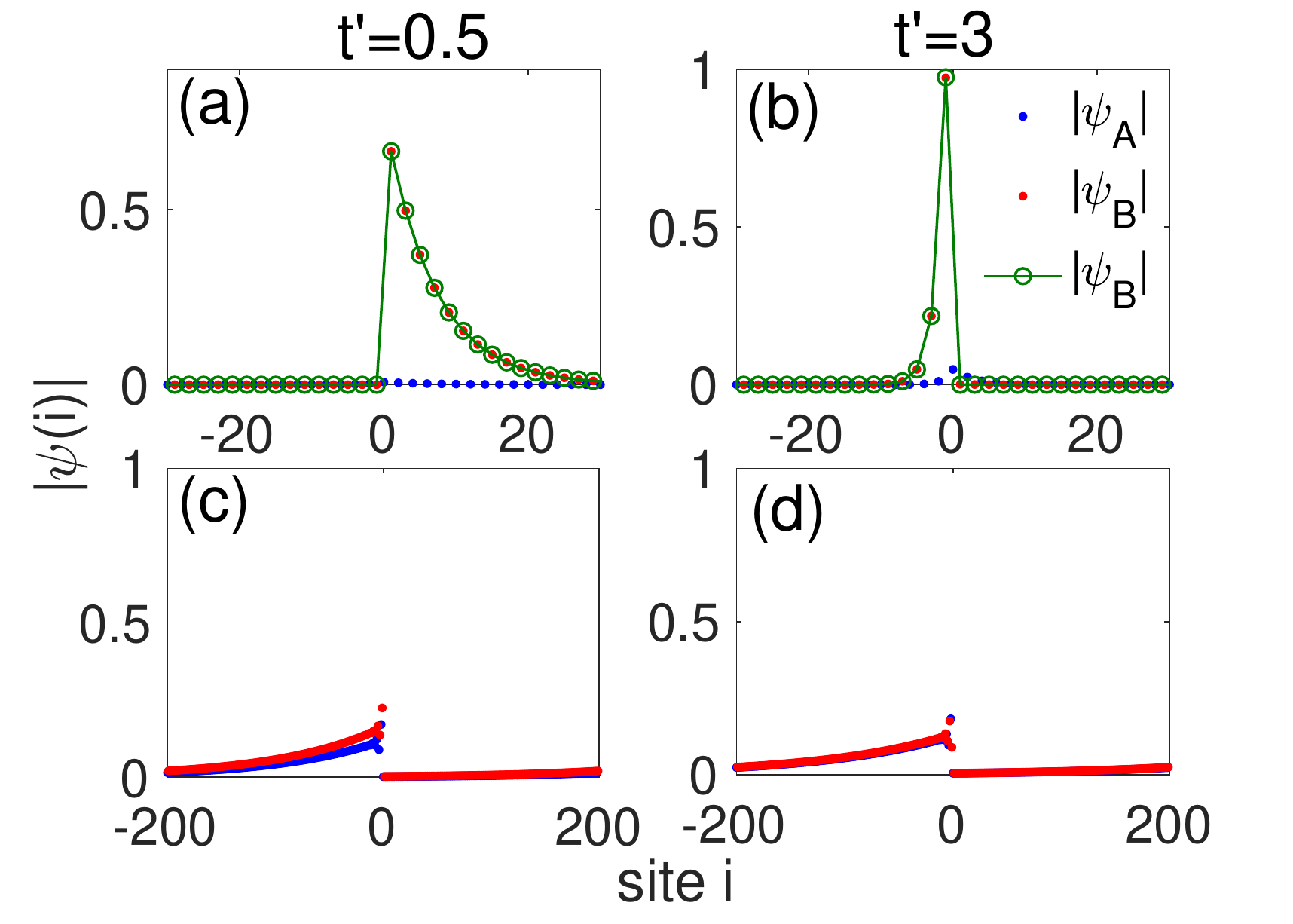}
\caption{Distribution of the mid-gap bound state for the system with $v=60$, $g=0.4$, (a) $t^{\prime }=0.5$ and (b) $t^{\prime }=3$, respectively. The green lines correspond to the wavefunction  obtained in the limit of $v \rightarrow \infty$. (c) and (d) give the distribution of a continuous state for the same system with $t^{\prime }=0.5$ and $t^{\prime }=3$, respectively.}
\label{fig7}
\end{figure}

In Fig. \ref{fig7}(a) and (b), we display distributions of wavefunction of the mid-gap bound state
for systems with $v=60$, $g=0.4$, $t^{\prime }=0.5$ and $t^{\prime }=3$, respectively. The numerical results for $v=60$ are shown to be almost indiscernible with the analytical results in the infinite $v$ limit.  It is clear that the mid-gap bound state locates on different sides of the impurity for the system with $t^{\prime }>1$ or $t^{\prime }<1$. Therefore, we can distinguish phases corresponding to $\nu_s=1$ from the phase  $\nu_s=0$ from the different distributions of the mid-gap state.
As a comparison, we also demonstrate the distribution of a
continuous state in Fig.\ref{fig7}(c) and (d) for the same system with $v=60$, $g=0.4$,  $t^{\prime }=0.5$ and $t^{\prime }=3$, respectively. The continuous state distributes asymmetrically and decays slowly from the impurity site. If we change the sign of $g$, the corresponding continuous states decay from other side of impurity. On the other hand, the distribution of mid-gap state on the left or right side of impurity only depends on the corresponding bulk system is in the phase of $\nu_s=1$ ($|t^{\prime }|>1$) or $0$  ($|t^{\prime }|<1$), and is not sensitive to the sign of $g$. More distributions of wavefunctions for the system with various parameters can be found in the appendix.

\section{Summary}
In summary, we have studied the single impurity problem in non-reciprocal topological lattices
and given the phase diagram of localized bound states.  We firstly solve the single impurity problem in a one-band non-reciprocal chain and give the analytical solution of bound state.
Due to the interplay of non-reciprocal
hopping and impurity potential, the localized bound state occurs only when
the potential strength exceeds a critical value. The influence of
non-reciprocal hoppings on the wavefuntions of both continuous states and
bound state are unveiled.
Then by mapping
the single-impurity problem in the non-reciprocal SSH model into a single impurity problem in a simple 1D non-reciprocal
lattice, we analytically determine the phase diagram and wavefunctions of the localized
bound states. In the large impurity strength limit, the
eigenvalue of the mid-gap bound state approaches zero and the boundary of
zero-mode bound state is found to be identical to the phase boundary of bulk
topological phase diagram. The distribution of mid-gap state can also discern topologically different bulk system. The existence of correspondence between impurity
zero-mode states and bulk phases suggests that we can detect the bulk phase
diagram from behavior of mid-gap impurity modes.

\begin{acknowledgments}
The work is supported by NSFC under Grants No.11974413 and the National Key
Research and Development Program of China (2016YFA0300600 and
2016YFA0302104).
\end{acknowledgments}

\appendix
\section{Distribution of wavefunctions for the non-reciprocal SSH model with a single impurity}
\begin{figure}[tbp]
\includegraphics[width=0.5\textwidth]{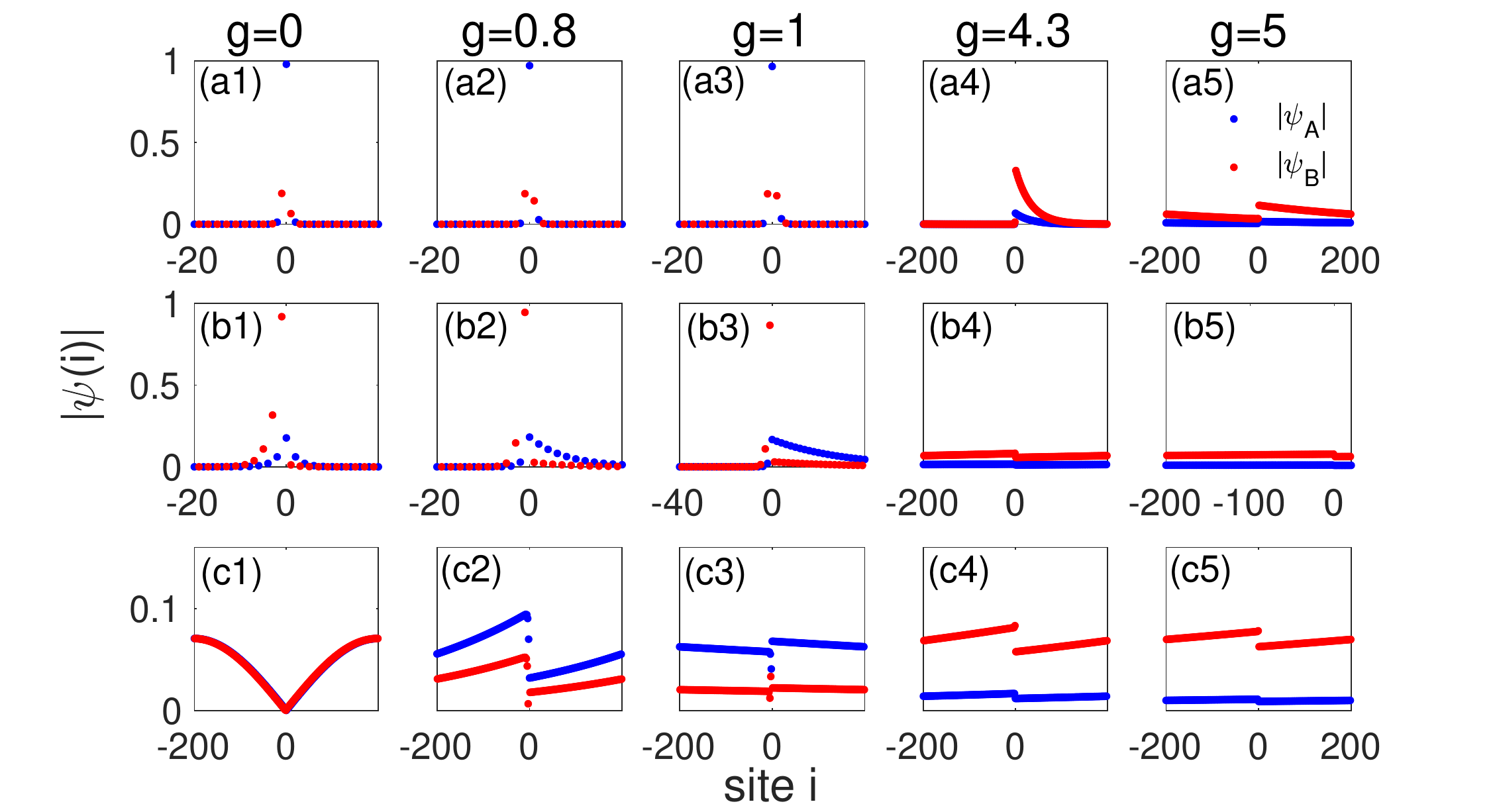}
\caption{The distributions of wavefunctions for the impurity model in the non-reciprocal SSH lattice with $%
t^{\prime }=3$ and $v=15$. The columns represent results for systems with
$g=0$, $0.8$, $1$, $4.3$ and $5$, respectively. (a1)-(a5) The wave functions correspond to eigenenergy $\varepsilon_{+}$. (b1)-(b5) The wave functions  correspond to eigenenergy $\varepsilon_{-}$. (c1)-(c5) The wave functions for a randomly chosen extended states.}
\label{fig8}
\end{figure}

In this appendix, we display the distribution of wavefunctions for the system of single impurity in the non-reciprocal SSH lattice with various parameters.
In Fig. \ref{fig8}(a) and (b), we display distributions of wavefunctions
corresponding to the two bound states for systems with $t^{\prime }=3$, $%
v^{\prime }=15$ and various $g$. When $g=0$, from the expressions of
the bound states in the main text, we can find that the wavefunction on sites $A$ distributes symmetrically
around $n=0$ and decays exponentially, but on sites $B$
distributes asymmetrically, due to the different multipliers of $\psi_B(n)$
in the regions $n>0$ and $n<0$. When $g\neq 0$, the wavefunction
 $\psi _{A}(n)$ distributes asymmetrically and decays
more quickly in one of sides of the impurity than in the other side, as
shown in Fig.\ref{fig8}(a) and (b). The bound state consists of two
exponentially decaying distributions on sites $A$ and $B$, respectively. When $g$
exceeds the first critical value ($g_{c1}\approx 1.1$), the mid-gap bound
state vanishes. Further increasing $g$ and exceeding the second critical
value ($g_{c2}\approx 4.36$), the bound state corresponding to $\varepsilon
_{+}$ also vanishes. As a comparison, we demonstrate the distribution of a
continuous state in Fig.\ref{fig8}(c) and find no obvious changes when $g$
exceeds both critical values. The distributions of wavefunctions are
consistent with the results of IPR shown in Fig.2(c) of the main text.
Distributions of bound-state wavefunctions for systems in the region of $%
t^{\prime }<1$ are found to display similar behaviors.
\begin{figure}[tbp]
\includegraphics[width=0.5\textwidth]{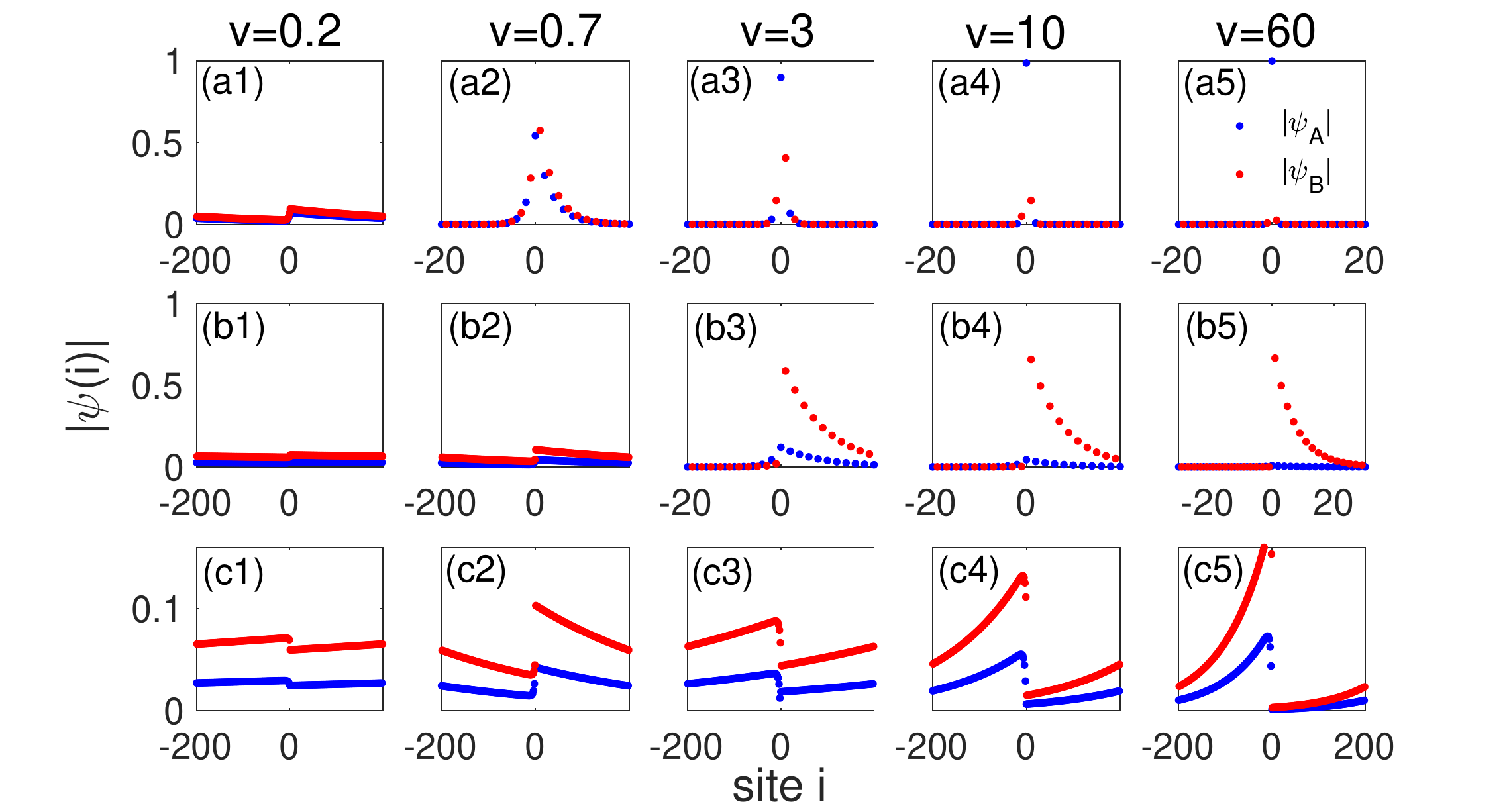}
\caption{The distributions of wavefunctions for the impurity model in the non-reciprocal SSH lattice with $%
t^{\prime }=0.5$ and $g=0.4$. The columns represent results for systems with
$v=0.2$, $0.7$, $3$, $10$ and $60$, respectively. (a1)-(a5) The wave functions correspond to eigenenergy
$\varepsilon_{+}$. (b1)-(b5) The wave functions oorrespond to eigenenergy $\varepsilon_{-}$. (c1)-(c5) The wave functions for a randomly chosen extended states.}
\label{fig9}
\end{figure}

\begin{figure}[tbp]
\includegraphics[width=0.5\textwidth]{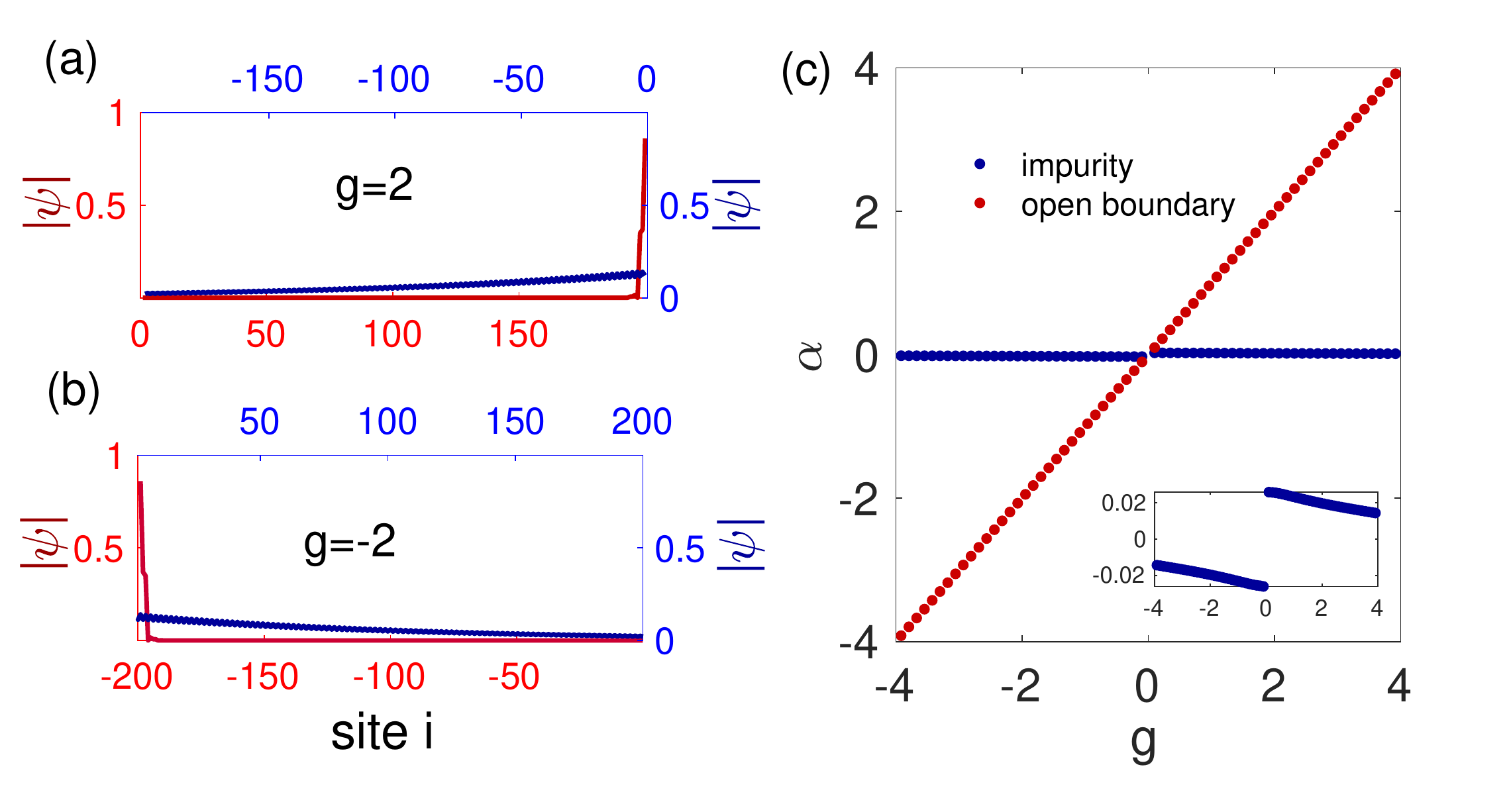}
\caption{ The distribution of the continuous state for impurity model (blue solid line)
and open-boundary model (red solid line) with $t'=3$, (a) $g=2$ and (b) $g=-2$, respectively. The parameter $v=200$ for the impurity model
and the site number is $400$. (c) The exponential decay constant for the distribution of
wavefunction of non-reciprocal SSH model with open boundary condition (red dots)
and with an impurity (blue dots).}
\label{fig10}
\end{figure}

Fig.\ref{fig9}(a1)-(a5) show the distributions of wavefunctions
corresponding to the eigenenergy $\varepsilon_+$ with $t^{\prime }=0.5$, $g=0.4$ and various $v$.
The formation of the bound states requires $f\left( \varepsilon_{+}v/t^{\prime }\right) <e^{-\left\vert g\right\vert }$, otherwise there is no bound state. For $v=0.2$, no bound state exists
and the corresponding state is a continuous state [See Fig. \ref{fig9} (a1)]. When $v$ exceeds a critical value, a bound state appears [See Figs.\ref{fig9} (a2)-(a5)]. The probability of distributions on B sites ($\psi_B$) decreases with the increase of $v$. In the limit of $v\rightarrow \infty$, $\psi_B$ is entirely suppressed and only $\psi_A$ exists. The wavefunctions for systems with $t^{\prime }>1$ are found to display similar behaviors.
Figs.\ref{fig9}(b1)-(b5) show the distributions of wavefunctions
corresponding to the eigenenergy $\varepsilon_-$  with $t^{\prime }=0.5$, $g=0.4$ and various $v$.
The formation of the mid-gap bound states requires $f\left( \varepsilon_{-}v/t^{\prime }\right) <e^{-\left\vert g\right\vert }$.  For $v=0.2$ and $0.7$, no mid-gap bound state exists and the corresponding state is a continuous state, as shown in Fig. \ref{fig9} (b1) and (b2). When $v$ exceeds the second critical value, a mid-gap bound state appears as shown in Fig.\ref{fig9} (b3)-(b5). The probability of distributions on A sites ($\psi_A$) decreases with the increase of $v$. In the limit of $v\rightarrow \infty$, $\psi_A$ is completely suppressed and the wavefunction only distributes on B sites with $\psi_B$ localized at the right side of the impurity.
The extended state may distribute on either the right or left region of the impurity for a given $g$ and small $v$, as shown in  Fig.\ref{fig9}(a1)-(c1).
In the large $v$ limit, the extended state for $g>0$ tends to distribute on the left region of the impurity, as shown in Fig.\ref{fig9}(c5).

For the large $v$ case, the distribution tendency of the continuous state is solely determined
by $g>0$ or $g<0$, which is similar to the skin effect in the open boundary case. However, the distribution for the impurity model decays very slowly and distribute in the whole region of the lattice, which is very different from the skin effect. Here we would like to compare their difference quantitatively.
The non-reciprocal SSH model under the open boundary condition can be mapped to a Hermitian SSH model by
a similarity transformation, $\mathcal{H}^{\prime}_{0}=U^{-1}\mathcal{H}_{0}U$,
where $\mathcal{H}^{\prime}_{0}$ is a standard SSH model ,i.e., the Hamiltonian $\mathcal{H}_{0}$ with $g=0$.
Here $U$ can be taken as a diagonal matrix whose diagonal elements are $\{1,e^{g},e^{g},e^{2g}
,e^{2g},\cdots,e^{Ng}\}$. The wavefunction of non-reciprocal SSH model can be written as $\left\vert \psi \right\rangle = U\left\vert\psi^{\prime}\right\rangle$,
where $\left\vert \psi ^{\prime}\right\rangle$ satisfies $\mathcal{H}^{\prime}_{0}\left\vert \psi^{\prime}\right\rangle=E\left\vert \psi^{\prime}\right\rangle$. As $\left\vert \psi ^{\prime}\right\rangle$ is always uniformly distributed except of the zero-mode state,
the 'non-Hermitain skin effect' can be understood as a result of the transformation $U$. When the $g<0$ ($g>0$), the eigenstates are
localized at the left (right) edge of the chain, as shown in Fig.\ref{fig10} (b) and (a) (red solid lines). In Fig .\ref{fig10}(c), we use red dots to represent the exponential decay constant $\alpha= \ln{(|\psi_A(n)|/|\psi_A(n-1)|)}\simeq g$ for the skin state of the system with OBC.
For the impurity system with large $v$, the distributions of continuous states decay exponentially but very slowly from the right or left
impurity site and the direction depends on the sign of $g$, as shown in Fig.\ref{fig10}(a) and (b) by the blue solid lines. In Fig. \ref{fig10}(c), we use blue dots to denote the exponential decay constant
$\alpha= \ln{(|\psi_A(n)|/|\psi_A(n-1)|)}$ for the impurity model, which shows the continuous states decaying much more slowly than the skin stats under open boundary condition.
The exponential decay constant $\alpha=0.0175$ for $g=2$ and $\alpha=-0.0175$
for $g=-2$.

\end{document}